\documentclass[twocolumn,showpacs,preprintnumbers,amsmath,amssymb,superscriptaddress,prd]{revtex4}

\usepackage{graphicx}
\usepackage{epstopdf}
\usepackage[T1]{fontenc}
\usepackage{dcolumn}
\usepackage{bm}

\def \figabbr{Fig.\ }
\def \eqnabbr{Eq.\ }
\def \secabbr{Section\ }

\begin{document}

\title{Gravitational Depolarization of Ultracold Neutrons: Comparison with Data}

\author{S.~Afach}
\affiliation{Paul Scherrer Institut, CH-5232 Villigen PSI,
Switzerland}
\affiliation{ETH Z\"{u}rich, Institute for Particle Physics, CH-8093 Z\"{u}rich, Switzerland}
\affiliation{Hans Berger Department of Neurology, Jena University Hospital, D-07747 Jena, Germany}

\author{N.J.~Ayres}
\affiliation{Department of Physics and Astronomy, University of Sussex, Falmer, Brighton BN1 9QH, UK}

\author{C.A.~Baker}\affiliation{STFC Rutherford Appleton Laboratory, Harwell, Didcot, Oxon OX11 0QX, UK}

\author{G.~Ban}
\affiliation{LPC Caen, ENSICAEN, Universit\'{e} de Caen, CNRS/IN2P3, Caen, France}

\author{G.~Bison}
\affiliation{Paul Scherrer Institut, CH-5232 Villigen PSI, Switzerland}

\author{K.~Bodek}
\affiliation{Marian Smoluchowski Institute of Physics, Jagiellonian University, 30-059 Cracow, Poland}





\author{M.~Fertl}
\altaffiliation{Present address: University of Washington, Seattle, United States of America}
\affiliation{Paul Scherrer Institut, CH-5232 Villigen PSI, Switzerland}

\author{B.~Franke}
\altaffiliation{Present address: Max-Planck-Institute of Quantum Optics, Garching, Germany.}
\affiliation{Paul Scherrer Institut, CH-5232 Villigen PSI, Switzerland}
\affiliation{ETH Z\"{u}rich, Institute for Particle Physics, CH-8093 Z\"{u}rich, Switzerland}

\author{P.~Geltenbort}\affiliation{Institut Laue-Langevin, BP 156, F-38042 Grenoble Cedex 9, France}

\author{K.~Green}\affiliation{STFC Rutherford Appleton Laboratory, Harwell, Didcot, Oxon OX11 0QX, UK}

\author{W.C.~Griffith}
\affiliation{Department of Physics and Astronomy, University of Sussex, Falmer, Brighton BN1 9QH, UK}

\author{M.~\surname{van~der~Grinten}}\affiliation{STFC Rutherford Appleton Laboratory, Harwell, Didcot, Oxon OX11 0QX, UK}

\author{Z.D.~Gruji\'c}
\affiliation{Physics Department, University of Fribourg, CH-1700 Fribourg, Switzerland}

\author{P.G.~Harris}
\email{p.g.harris@sussex.ac.uk}\affiliation{Department of Physics and Astronomy, University of Sussex, Falmer, Brighton BN1 9QH, UK}

\author{W.~Heil}
\affiliation{Institut f\"{u}r Physik, Johannes-Gutenberg-Universit\"{a}t, D-55128 Mainz, Germany}

\author{V.~H\'{e}laine}
\affiliation{LPC Caen, ENSICAEN, Universit\'{e} de Caen, CNRS/IN2P3, Caen, France}

\author{P.~Iaydjiev}\altaffiliation{On leave of absence from Institute of Nuclear Research and Nuclear Energy, Sofia, Bulgaria}\affiliation{STFC Rutherford Appleton Laboratory, Harwell, Didcot, Oxon OX11 0QX, UK}

\author{S.N.~Ivanov}\altaffiliation{On leave of absence from Petersburg Nuclear Physics Institute, Russia}\affiliation{STFC Rutherford Appleton Laboratory, Harwell, Didcot, Oxon OX11 0QX, UK}

\author{M.~Kasprzak}
\altaffiliation{Present address: Instituut voor Kern- en Stralingsfysica, Katholieke Universiteit Leuven, B-3001 Leuven, Belgium}
\affiliation{Physics Department, University of Fribourg, CH-1700 Fribourg, Switzerland}

\author{Y.~Kermaidic}
\affiliation{LPSC, Universit\'{e} Grenoble Alpes, CNRS/IN2P3, Grenoble, France}

\author{K.~Kirch}
\affiliation{Paul Scherrer Institut, CH-5232 Villigen PSI, Switzerland}
\affiliation{ETH Z\"{u}rich, Institute for Particle Physics, CH-8093 Z\"{u}rich, Switzerland}


\author{H.-C.~Koch}
\affiliation{Physics Department, University of Fribourg, CH-1700 Fribourg, Switzerland}
\affiliation{Institut f\"{u}r Physik, Johannes-Gutenberg-Universit\"{a}t, D-55128 Mainz, Germany}

\author{S.~Komposch}
\affiliation{Paul Scherrer Institut, CH-5232 Villigen PSI, Switzerland}
\affiliation{ETH Z\"{u}rich, Institute for Particle Physics, CH-8093 Z\"{u}rich, Switzerland}

\author{A.~Kozela}
\affiliation{Henryk Niedwodniczanski Institute for Nuclear Physics, Cracow, Poland}

\author{J.~Krempel}
\affiliation{ETH Z\"{u}rich, Institute for Particle Physics, CH-8093 Z\"{u}rich, Switzerland}

\author{B.~Lauss}
\affiliation{Paul Scherrer Institut, CH-5232 Villigen PSI, Switzerland}

\author{T.~Lefort}
\affiliation{LPC Caen, ENSICAEN, Universit\'{e} de Caen, CNRS/IN2P3, Caen, France}

\author{Y.~Lemi\`{e}re}
\affiliation{LPC Caen, ENSICAEN, Universit\'{e} de Caen, CNRS/IN2P3, Caen, France}


\author{M.~Musgrave}
\affiliation{Department of Physics and Astronomy, University of Sussex, Falmer, Brighton BN1 9QH, UK}

\author{O.~Naviliat-Cuncic} \altaffiliation{Present address: Michigan State University, East-Lansing, USA.}
\affiliation{LPC Caen, ENSICAEN, Universit\'{e} de Caen, CNRS/IN2P3, Caen, France}

\author{J.M.~Pendlebury}
\affiliation{Department of Physics and Astronomy, University of Sussex, Falmer, Brighton BN1 9QH, UK}

\author{F.M.~Piegsa}
\affiliation{ETH Z\"{u}rich, Institute for Particle Physics, CH-8093 Z\"{u}rich, Switzerland}

\author{G.~Pignol}
\affiliation{LPSC, Universit\'{e} Grenoble Alpes, CNRS/IN2P3, Grenoble, France}

\author{C.~Plonka-Spehr}
\affiliation{Institut f\"{u}r Kernchemie, Johannes-Gutenberg-Universit\"{a}t, Mainz, Germany}

\author{P.N.~Prashanth}
\affiliation{Instituut voor Kern- en Stralingsfysica, Katholieke Universiteit Leuven, B-3001 Leuven, Belgium}

\author{G.~Qu\'{e}m\'{e}ner}
\affiliation{LPC Caen, ENSICAEN, Universit\'{e} de Caen, CNRS/IN2P3, Caen, France}

\author{M.~Rawlik}
\affiliation{ETH Z\"{u}rich, Institute for Particle Physics, CH-8093 Z\"{u}rich, Switzerland}

\author{D.~Rebreyend}
\affiliation{LPSC, Universit\'{e} Grenoble Alpes, CNRS/IN2P3, Grenoble, France}

\author{D.~Ries}
\affiliation{Paul Scherrer Institut, CH-5232 Villigen PSI, Switzerland}
\affiliation{ETH Z\"{u}rich, Institute for Particle Physics, CH-8093 Z\"{u}rich, Switzerland}

\author{S.~Roccia}
\affiliation{CSNSM, Universit\'{e} Paris Sud, CNRS/IN2P3, Orsay, France}

\author{D.~Rozpedzik}
\affiliation{Marian Smoluchowski Institute of Physics, Jagiellonian University, 30-059 Cracow, Poland}


\author{P.~Schmidt-Wellenburg}
 \affiliation{Paul Scherrer Institut, CH-5232 Villigen PSI, Switzerland}

\author{N.~Severijns}
\affiliation{Instituut voor Kern- en Stralingsfysica, Katholieke Universiteit Leuven, B-3001 Leuven, Belgium}

\author{D.~Shiers}
\affiliation{University of Sussex, Falmer, Brighton BN1 9QH, UK}

\author{J.A.~Thorne}
\affiliation{Department of Physics and Astronomy, University of Sussex, Falmer, Brighton BN1 9QH, UK}


\author{A.~Weis}
\affiliation{Physics Department, University of Fribourg, CH-1700 Fribourg, Switzerland}

\author{E.~Wursten}
\affiliation{Instituut voor Kern- en Stralingsfysica, Katholieke Universiteit Leuven, B-3001 Leuven, Belgium}


\author{J.~Zejma}
\affiliation{Marian Smoluchowski Institute of Physics, Jagiellonian University, 30-059 Cracow, Poland}

\author{J.~Zenner}
\affiliation{Paul Scherrer Institut, CH-5232 Villigen PSI, Switzerland}
\affiliation{ETH Z\"{u}rich, Institute for Particle Physics, CH-8093 Z\"{u}rich, Switzerland}
\affiliation{Institut f\"{u}r Kernchemie, Johannes-Gutenberg-Universit\"{a}t, Mainz, Germany}

\author{G.~Zsigmond}
\affiliation{Paul Scherrer Institut, CH-5232 Villigen PSI, Switzerland}

\date{\today}

\begin{abstract}

We compare the expected effects of so-called gravitationally enhanced depolarization of ultracold neutrons to measurements carried out in a spin-precession chamber exposed to a variety of vertical magnetic-field gradients.  In particular, we have investigated the dependence upon these field gradients of spin depolarization rates and also of shifts in the measured neutron Larmor precession frequency.   
We find excellent qualitative agreement, with gravitationally enhanced depolarization accounting for several previously unexplained features in the data.

\end{abstract}

\maketitle

\section{Introduction}
\label{sec:intro}

Ultracold neutrons (UCN) are neutrons of extremely low energy, typically ~200\,neV or less, which can be stored in material bottles and which are routinely used in experiments such as the ongoing search for the neutron electric dipole moment (nEDM).  Collisions with the containing walls are elastic, so the UCN never thermalize.   Being of such low energy, they ``sag'' under gravity, and rather than being distributed uniformly throughout their storage vessel their density decreases with increasing height, with each specific energy group having its own center of mass.    In the presence of a vertical magnetic-field gradient, the average magnetic field sampled by the neutrons will therefore depend upon the neutron energy.  The implications of this stratification have been discussed in earlier work \cite{Harris_2014, knecht09}, but, in summary, it results in a relative dephasing of the neutrons in different energy bins, which then alters the measured Larmor spin-precession frequency.  This phenomenon is referred to as {\em gravitationally enhanced depolarization}, in contrast to the {\em intrinsic} depolarization  that takes place within each energy bin as a result of the neutrons sampling different fields as they move around the storage volume.
A key distinction is the asymmetric nature of the gravitationally induced dephasing, as shown in \figabbr 3 of \cite{Harris_2014}, with the lowest-energy neutrons playing a particularly crucial role.  The resulting nonlinearities in frequency response as a function of applied magnetic field gradients represent potential sources of systematic uncertainty in precision experiments such as nEDM searches \cite{Pendlebury_2004,Baker_2006,afach15c}.  Since such experiments provide tight constraints on physics beyond the Standard Model, with consequent implications for particle theory and cosmology, a full understanding of the phenomenon is essential.

In this article, we compare our experimentally measured results, both in terms of frequency shifts and of depolarization rates, with those anticipated from theoretical calculations.  
 We begin in \secabbr \ref{sec:spectra} with a discussion of the spectrum of UCN within our storage cell; this underlies the subsequent calculations of the gravitationally enhanced depolarization.  We give an overview of the calculations themselves in \secabbr \ref{sec:calcs}.  In \secabbr \ref{sec:intrinsic} we discuss the basic intrinsic-depolarization mechanisms, which are revealed to make only a minor contribution to the frequency shifts. We then present, in \secabbr \ref{sec:alpha_peak}, a direct comparison of the anticipated and measured polarization $\alpha$ remaining after 180\,s of storage in a range of applied $B$-field gradients.  In \secabbr \ref{sec:R_curve_large} and \ref{sec:R_curve_closeup}, we consider the frequency shifts that arise from this phenomenon, before finally discussing in \secabbr \ref{sec:edm_implications} the possible implications for nEDM experiments, including the current world limit in particular.

The measurements described in this article form part of a program of work  \cite{Baker_2011} aimed at an accurate determination of the nEDM, currently being carried out at the new high-intensity UCN source \cite{Lauss_2014} based at the Paul Scherrer Institute (PSI).  The experimental apparatus and procedures are described in substantial detail in \cite{Afach_2014}.  The apparatus is based upon that used \cite{Baker_2014} in an earlier nEDM measurement at the Institut Laue-Langevin (ILL) \cite{Baker_2006}, but substantially upgraded with the incorporation, in particular, of an array of Cs magnetometers \cite{Knowles_2009}, a system for the simultaneous detection of both neutron spin states \cite{afach15}, and a set of active compensation coils that provide dynamic shielding of external magnetic fields \cite{Afach_2014b}.

The 1\,$\mu$T magnetic holding field $B_0$ within the EDM spectrometer is primarily vertical, so $B_0 \approx B_z$, although there are small transverse components $B_x, B_y$ present at the $\sim$ few nT level.  We define
\begin{equation}
B_0 = B_z + \frac{1}{2}\frac{B_t^2}{B_z},
\end{equation}
where $B_t^2 = B_x^2 + B_y^2$.

In order to compensate for changes in $B_0$, a cohabiting atomic mercury magnetometer \cite{Green_1998} is used to make precise real-time measurements of the volume-averaged field within the UCN storage cell. Under an applied vertical magnetic-field gradient $\partial B_z/\partial z$,  the measured ratio $R$ of neutron to mercury precession frequencies undergoes a relative change of, to first order, 
\begin{equation}
\label{eqn:deltaR}
\frac{\delta R}{R} = \frac{1}{B_0}\frac{\partial B_0}{\partial z} \Delta h, 
\end{equation}
where $\Delta h$ is the difference between the centers of mass of the populations of (thermal) mercury atoms and (ultracold) neutrons.   Precise measurements of this frequency-ratio dependence are the subject of \cite{Afach_2014}.  As we shall see, gravitationally enhanced depolarization can impose a substantial nonlinearity in this relationship: indeed, we are unaware of any other mechanism that can do so to the extent required to match our observations.

\section{Input spectra}
\label{sec:spectra}

The extent of gravitational depolarization clearly depends heavily upon the spectrum of stored UCN.  We have recently carried out a series of  measurements  using a spin-echo technique \cite{afach15b},  from which we were able to derive the distribution of energies of UCN remaining after 220\,s of storage in our apparatus.  The resulting fitted spectrum is parameterized by
\begin{equation}
\label{eq:spectrum} 
		p(E) = A\cdot E^{1/2}\cdot\frac{1}{1+e^{\frac{E_0-E}{\Delta E_0}}}\cdot 
		\frac{1}{1+e^{\frac{E-E_1}{\Delta E_1}}},
\end{equation}
where  $A$ is an arbitrary normalization, $E_0$ = 
7.7\,neV, $ \Delta E_0= 1$\,neV,  $E_1 = $ 
28.7\,neV, and  $\Delta E_1 =6.25 $\,neV.  The form of this parameterization is based on a very general distribution $n(E)\text{d}E \propto E^{1/2}\text{d}E$ from the low-energy tail of a Maxwell-Boltzmann distribution, allowing for low- and high-energy cut-offs.

  The spin-echo technique is particularly sensitive to the presence of low-energy UCN, but once the neutrons start to populate the bottle more or less uniformly it becomes increasingly difficult to distinguish between different energies.  This is clear from Figs.\ 2(a) and 2(b) in \cite{afach15b}, where the low-energy tails are fitted well but the high-energy region produces less reliable results.  Furthermore, the spin-echo measurements were carried out at a storage time of 220\,s, whereas the polarization and frequency-ratio measurements used in the current analysis were carried out at a storage time of 180\,s.  On both counts, therefore, we should not be surprised if the actual spectrum were to be somewhat firmer than that arising from the spin-echo measurement.

We have also used the package MCUCN \cite{Bodek_2011} to carry out a detailed simulation of the UCN within our apparatus, which yields an alternative estimate of the spectrum after 180\,s of storage.  The simulation is based upon very detailed modeling of the PSI UCN source, beamline, and guides, as well as of the nEDM storage vessel.  The latter consists of aluminum electrodes coated with diamond-like carbon, which form the floor and roof, and between them an insulating cylindrical polystyrene ring coated with deuterated polystyrene to provide radial containment.    The simulation accounts for losses during storage both from $\beta$ decay and as a result of wall collisions \cite{golub_UCN_book}, with the ``loss-per-bounce'' factor $f = W/V$ (where $W$, $V$ are the imaginary and real parts, respectively, of the Fermi potential) set to a common value of $3\times10^{-4}$ for the electrodes and for the insulator walls.  In fact, although $V$ is well known, $W$ is difficult to determine.  Losses are likely to be dominated by hydrogen that has diffused into the containing surfaces, and can -- because it has an extremely high incoherent-scattering cross section -- substantially influence loss rates without significantly altering the surface potential $V$.   Using an average value of $f$ for all of the containing walls appears a reasonable approach, and the number here arises from earlier simulation-based studies \cite{kuzniak08} that were tuned to match the observed numbers of neutrons stored as a function of time.

Any damage to the coatings on either the electrodes or the insulating walls would result in an area of reduced Fermi potential that would preferentially deplete the higher-energy neutrons.  The same is true of small gaps, which may not be completely accurately modeled in the simulation.  The actual spectrum, therefore, is likely to be somewhat softer than the simulation would suggest.

\begin{figure} [ht]
\begin{center}
\resizebox{0.5\textwidth}{!}{
\includegraphics
{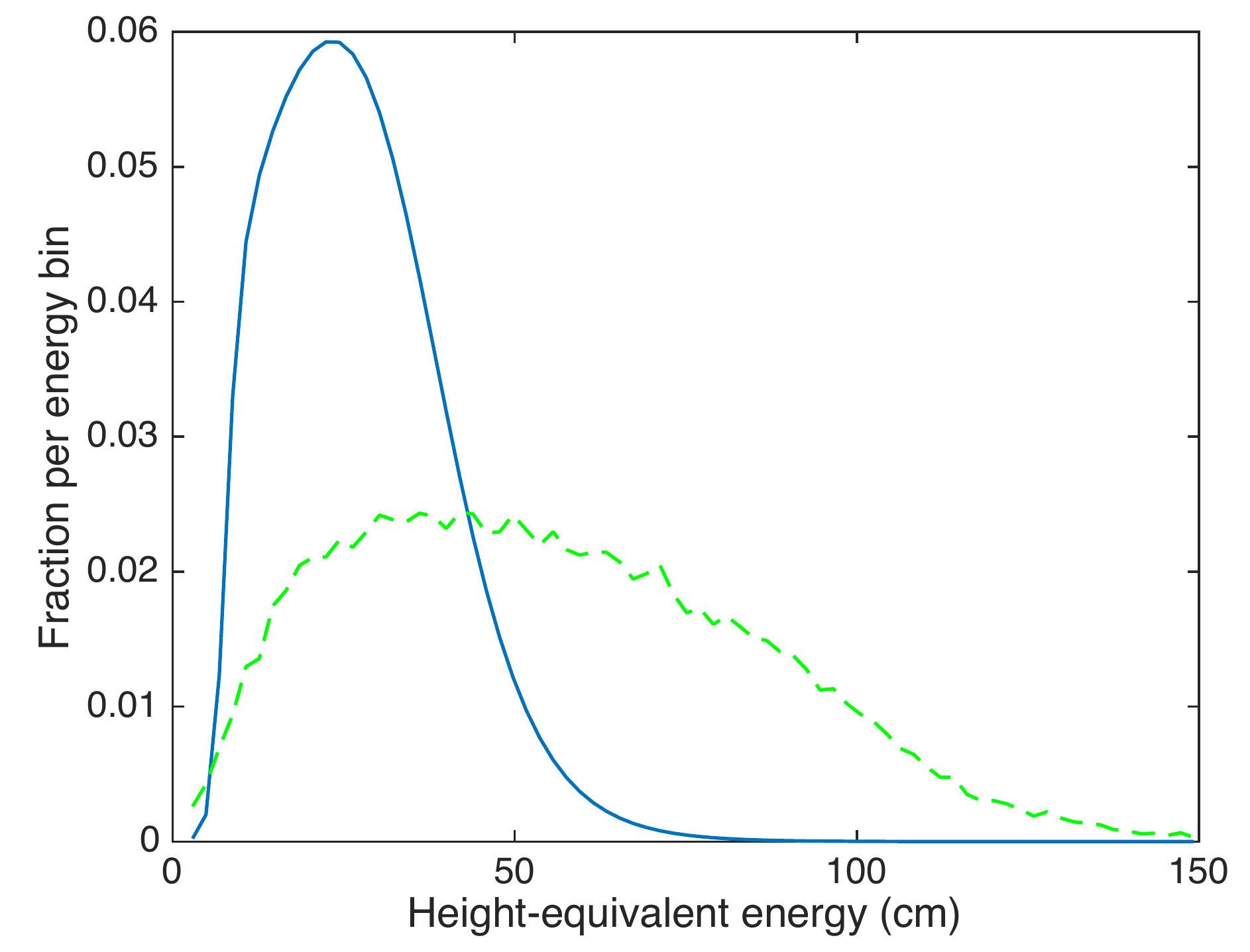}
}
\end{center}
\caption{(Color online) Estimated energy spectrum of stored UCN from  spin-echo measurement (blue solid line) and from simulation (green dashed line).  Each  is normalized to give unit total area under the curve.  }
\label{fig:spectra}
\end{figure}

The spectra resulting from the Monte Carlo (MC) simulation and spin-echo (SE) studies are shown in \figabbr \ref{fig:spectra}.  It is useful to refer to UCN energies $E$ in terms of the maximum height $\epsilon = E/(mg)$ attainable under gravity in a trap with no vertical confinement, and to this end the abscissa is in units of cm. 

We note here for clarity that the energies are defined to be the kinetic energies that the UCN would have at the floor of the storage vessel, i.e.\ at the bottom electrode.  


\section{Calculations}
\label{sec:calcs}

Calculations of the gravitationally enhanced depolarization effect are relatively straightforward to carry out.  Phase-space arguments \cite{Pendlebury_1994} show that the variation of density $\rho$ with height $z$ of UCN of height-equivalent energy $\epsilon$ is (for $z < \epsilon$) given by
\begin{equation}
\label{eqn:height_dist}
\rho(z,\epsilon) =\rho(0,\epsilon) \left(1-\frac{z}{\epsilon}\right)^{1/2},
\end{equation}
assuming sufficiently diffuse reflections for the phase space to be approximately uniformly populated on a timescale short compared to the storage time.  (Our Monte Carlo simulations confirm that this typically takes place within 5 s of closing the UCN shutter if more than 10\% of reflections are diffuse, and more quickly still with higher diffusivity and also for the more energetic of the neutrons in our spectrum.)    The height distribution of \eqnabbr \ref{eqn:height_dist} will then be reflected in the distribution of average magnetic fields to which UCN of any particular energy will be exposed, from which the distribution (appropriate to that UCN energy) of integrated phases acquired after 180\,s of free Larmor precession 
can be calculated.  This procedure is carried out for all UCN energies across the spectrum, accounting for the relative populations of each energy bin.  The resulting total array of integrated phases $\phi_i$ is then subject to a Ramsey-type analysis, where, as discussed in \cite{Harris_2014}, the net frequency is determined by the reference phase
\begin{equation}
\label{eqn:ref_phase}
\hat{\phi} = 2n\pi + \tan^{-1} \left(\frac{\langle \sin \phi_i \rangle}{\langle \cos \phi_i \rangle} \right),
\end{equation}
divided by the Ramsey coherence time (180\,s in  this case).  Note that the  $2n\pi $ term, which arises because the Ramsey technique measures phases modulo $2\pi$,  is relatively easily accounted for by, for example, monitoring the discrepancy between the reference phase and the mean of the array of time-integrated phases, and adding (or subtracting) factors of $2\pi$ as appropriate to compensate.  The nonlinearities in response referred to earlier primarily arise when the lowest-energy UCN, which do not reach the roof of the storage trap, have an integrated phase that differs by more than $\pi$ radians from the reference phase: they then ``wrap around'' and appear to enhance the high-energy tail of the distribution.  We will refer to this phenomenon as ``Ramsey wrapping''.

Effects due to intrinsic depolarization, arising from both vertical and horizontal field gradients, can also be included by appropriate weighting of the distribution of phases.  We discuss this in some detail in the following \secabbr.

\section{Intrinsic depolarization mechanisms}
\label{sec:intrinsic}

Detailed calculations of intrinsic depolarization within magnetic-field gradients have been carried out elsewhere \cite{Schearer_1965,Cates_1988,Cates_1988b, McGregor_1990,Schmid_2008,Lamoreaux_2005,Golub_2011,Pignol_2012}.  There are four relevant scenarios to consider: vertical gradients $\partial B_z/\partial z$; horizontal gradients of the form  $\partial B_z/\partial x$; transverse fields $B_t$  and their gradients; and wall collisions involving small magnetic impurities.   We present here some simple and rather intuitive models of the depolarization mechanisms, and we discuss possible implications for the polarization and frequency-ratio measurements.  Throughout this Section, where calculations are dependent upon an input spectrum we use that derived from the SE measurement.

\subsection{Vertical gradients: $\partial B_z/\partial z$}

Here we consider UCN confined within a vertical magnetic-field gradient $\partial B_z/\partial z$.  Let the confining trap be a cylinder of height $H$ and radius $r$.  Following \cite{Pendlebury_1994}, we replace $H$ with an ``effective height'' $\mathcal{H}(\epsilon)$ which is simply defined as the lesser of $H, \epsilon$; this accommodates UCN with energies too low to reach the roof of the trap.

The following method is based upon  that outlined in the derivation of \eqnabbr 68 in \cite{Kleppner_1962}.  We shall consider our trap to be divided by a horizontal plane into two halves, with average field strengths that each differ from the field at the center plane by $\Delta B_z = (\partial B_z/\partial z)\mathcal{H}/4$.  Let the average dwell time for UCN in each half of the trap be $t_w$. Using the standard kinetic-theory result (due to Clausius \cite{pendlebury_kinetic_theory}) that the rate of wall collisions per particle is $Av/(4V)$, where $A$ is here the area of the dividing plane,  $V$ is half of the containing volume and $ v $ is the speed of the particles, we can calculate the rate of passage between the two halves.   From this we find  
\begin{equation}
t_w = \frac{2\mathcal{H}}{v}.
\end{equation}

Consider now a single UCN.  Effectively, a coin is tossed once every $t_w$ to determine which side of the trap the neutron is in.  Over a storage time $t$, this decision is therefore made $N = t/t_w$ times.  The number of times $n$ for which the UCN is on the side with the stronger field is binomially distributed with mean $N/2$ and variance $N/4$.  The additional $\int B\cdot dt$ experienced by this UCN is $\sim 2(n-N/2) t_w \Delta B_z$ (where the factor 2 accounts for the fact that when it is not in the stronger-field region it is in the weaker-field region). Multiplying this by the neutron gyromagnetic ratio $\gamma_n$ gives the extra precession angle $\theta_t$ away from the mean.  The polarization is the average projection upon the mean precession vector, and using
\begin{equation}
e^{-t/T_2} \sim 1- \frac{t}{T_2}+ ...,
\end{equation}
where $T_2$ is the transverse spin-relaxation time, together with $\cos\theta \sim 1-\theta^2/2+...$, we find that
\begin{equation}
\label{eqn:T2vgi1}
\frac{t}{T_{2,\mathrm{vgi}}}\sim\frac{\langle \theta_t^2\rangle}{2}=2\left(\frac{N}{4}\right)\gamma_n^2\Delta B_z^2 t_w^2,
\end{equation}
where the subscript ``vgi'' stands for ``vertical gradient, intrinsic''.  This yields
\begin{equation}
\label{eqn:T2dBzdz}
T_{2,\mathrm{vgi}}\sim\frac{2}{\gamma_n^2 \Delta B_z^2 t_w} = \frac{16v}{\mathcal{H}^3\gamma_n^2 (\partial B_z/\partial z)^2}.
\end{equation}

\begin{figure} [ht]
\begin{center}
\resizebox{0.5\textwidth}{!}{
\includegraphics
{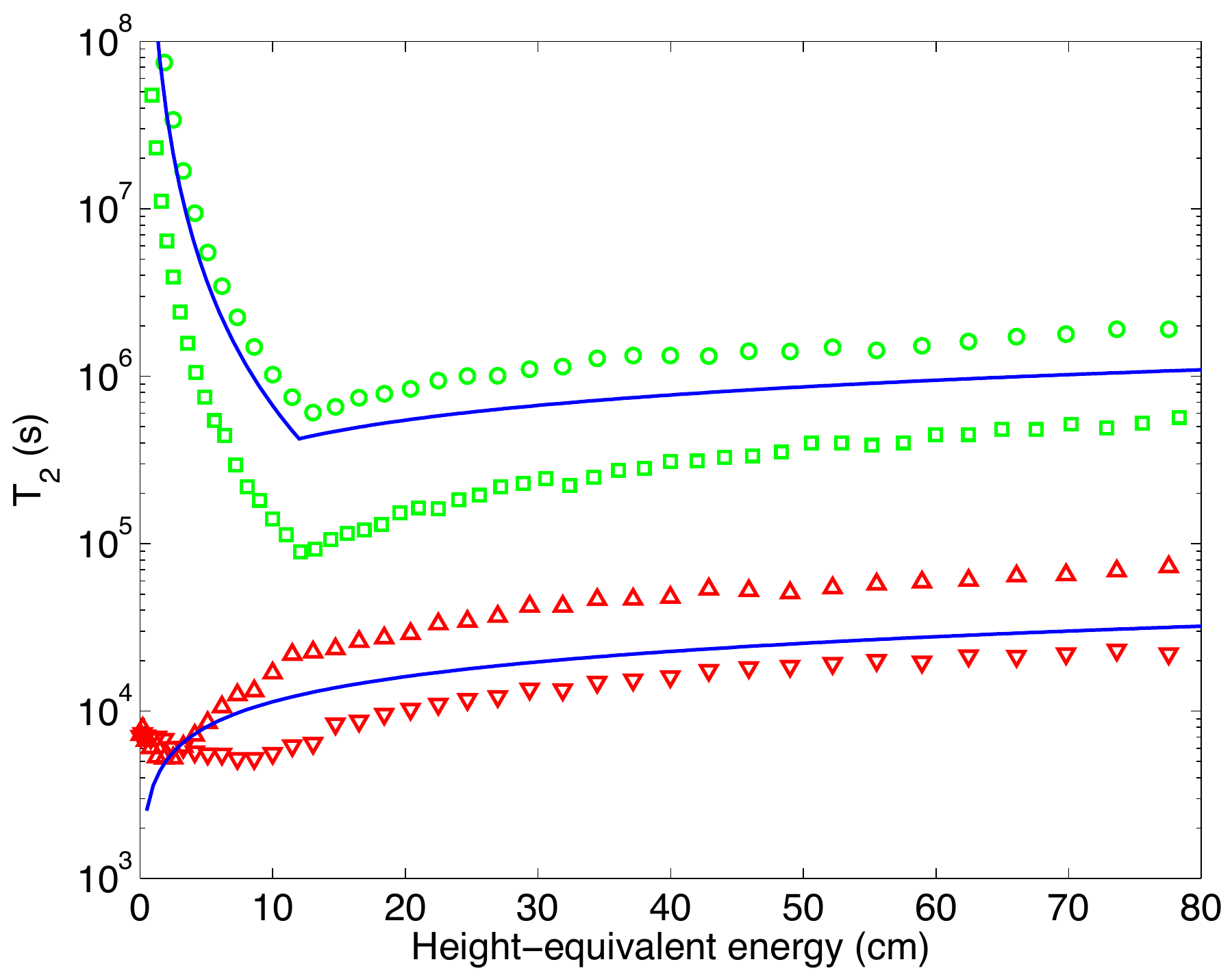}
}
\end{center}
\caption{(Color online) Intrinsic depolarization times $T_2$ for UCN in 10\,pT/cm magnetic-field gradients.  The upper set are for vertical gradients $\partial B_z/\partial z$ with the results of simulations for completely diffuse (green circles) and 80\% specular (green squares) reflections; the solid blue line is the analytical approximation of \eqnabbr \ref{eqn:T2dBzdz}.  The lower set are for transverse gradients in the vertical field, of the form $\partial B_z/\partial x$.  Red upwards-pointing (downwards-pointing) triangles are for diffuse (80\% specular) reflections, and the solid blue line represents the analytical approximation of \eqnabbr \ref{eqn:T2dBzdx}.  $T_2$ scales as the inverse square of the applied gradient. }
\label{fig:T2}
\end{figure}

The upper solid blue line in \figabbr \ref{fig:T2} shows the prediction of \eqnabbr \ref{eqn:T2dBzdz}, and despite the rather crude nature of its derivation it is seen to lie nicely between the results of our simulations for completely diffuse reflections (green circles) and for the case where the probability of specular reflections is 80\% (green squares).  We therefore use \eqnabbr \ref{eqn:T2dBzdz} in our calculations going forward, bearing in mind nonetheless that there is some uncertainty in the size of its contribution.  The simulated results also appear in \figabbr 2 of \cite{Harris_2014}, where it is shown that the expected dependence upon $(\partial B_z/\partial z)^2$ holds true over a wide range of gradients.

\begin{figure} [ht]
\begin{center}
\resizebox{0.5\textwidth}{!}{
\includegraphics
{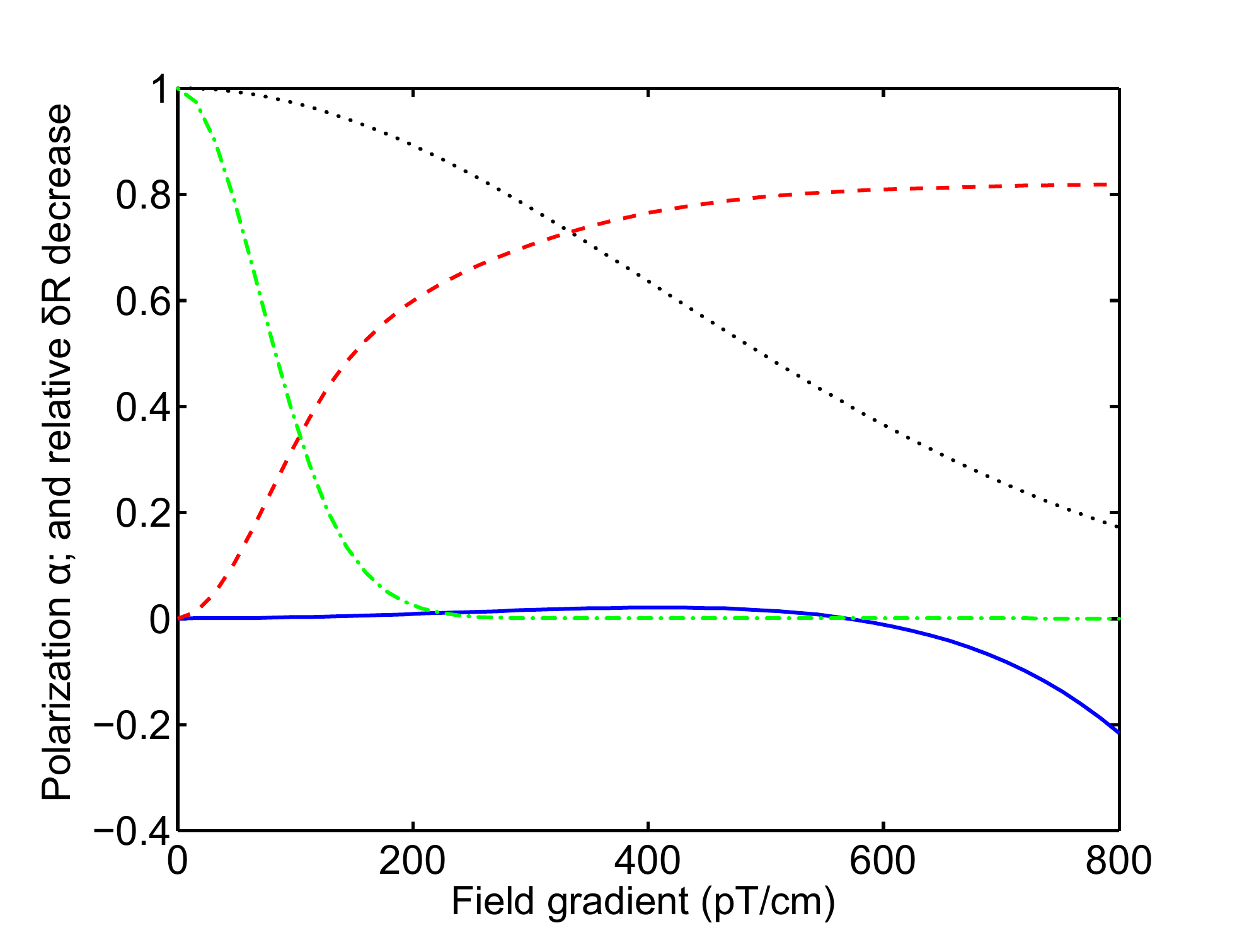}
}
\end{center}
\caption{(Color online) Polarization $\alpha$, and fractional decrease in the shift $\delta R$ in frequency ratio, arising from intrinsic depolarization only, after 180\,s of storage.  Black dotted (green dot-dashed): $\alpha$ from $\partial B_z/\partial z$ ($\partial B_z/\partial x$).    Blue solid (red dashed): relative decrease in $\delta R$, i.e.\ in the slope of $R$ vs.\ $\partial B_z/\partial z$, for vertical (horizontal) gradients in $B_z$. Typically, $\partial B_z/\partial x$ may be 30-50\,pT/cm.
}
\label{fig:deltah_shift}
\end{figure}

After a measurement time $t$, the polarization $\alpha$ is reduced by a factor 
\begin{equation}
e^{-{t}/{T_{2,\mathrm{vgi}}}}\sim\left(1-\frac{t\mathcal{H}^3\gamma_n^2}{16v}\left(\frac{\partial B_z}{\partial z}\right)^2+...\right),
\end{equation}
which implies a parabolic profile to the dependence of $\alpha$ upon the vertical magnetic-field gradient.  We see this in \figabbr \ref{fig:deltah_shift}, which shows (dotted black line) the spectrum-weighted average $\alpha$ as a function of the applied vertical gradient.

We can now proceed to make a rough estimate of the extent to which this intrinsic depolarization may result in a shift in the measured neutron frequency.  
 Intuitively, for example, it might seem that if low-energy UCN depolarize more quickly than their high-energy counterparts, they would have less of a role to play in determining the frequency (since the uncertainty on the frequency measurement is inversely proportional to the polarization $\alpha$).  In consequence, the frequency measurement may appear to arise from a somewhat stiffer spectrum than is actually the case,  thereby raising the effective center of mass of the neutrons and reducing the factor $\Delta h$ in \eqnabbr \ref{eqn:deltaR}.  In order to consider this, let us for the time being imagine that we can make absolute frequency measurements without the complication of the modulo 2$\pi$ arising from the Ramsey measurement, thus ignoring the Ramsey wrapping that is characteristic of the behaviour of the lowest-energy UCN \cite{Harris_2014}.  

By definition, the height difference $\Delta h$ between the centers of mass of mercury and UCN (\eqnabbr \ref{eqn:deltaR}) is
\begin{equation}
\Delta h = \hat{z}_\mathrm{Hg} - \frac{\iint z\rho(z,\epsilon) \,dz\, d\epsilon}{\iint \rho(z,\epsilon) \,dz\, d\epsilon},
\end{equation} 
where $\hat{z}_\mathrm{Hg}$ is the center of mass of the mercury atoms.  
We note in passing the standard result that, for $\epsilon \gg H$, $\Delta h = H^2/(24\epsilon)$.  When $\epsilon<H$, $\Delta h = H/2 - 0.4\epsilon$ instead; in intermediate regimes, $\Delta h$ may be derived from a more precise expression for the center of mass \cite{Harris_2014,afach15b}.  

We now define an effective height difference
\begin{equation}
\Delta h_\mathrm{eff} = \hat{z}_\mathrm{Hg} - \frac{\iint z \alpha(\epsilon)^2 \rho(z,\epsilon) \,dz\, d\epsilon}{\iint \alpha(\epsilon)^2 \rho(z,\epsilon) \,dz\, d\epsilon},
\end{equation} 
which takes into account the relative contribution of each energy bin to the frequency measurement.  The intrinsic-depolarization induced fractional decrease in the frequency ratio, away from that anticipated by \eqnabbr \ref{eqn:deltaR}, is then $\Delta(\delta R)/\delta R = (\Delta h-\Delta h_\mathrm{eff})/\Delta h$.  This function is shown (solid blue line)   in \figabbr \ref{fig:deltah_shift}.  We see that the effect stays at the 2\% level or below until quite large vertical gradients, in excess of 500\,pT/cm, by which time (as we shall see in \secabbr \ref{sec:R_curve_closeup} below) the Ramsey wrapping will in any case long since have taken hold.

\subsection{Horizontal gradients: $\partial B_z/\partial x$}

We now carry out exactly the same calculations for the case of horizontal changes $\partial B_z/\partial x$ in the vertical magnetic field.  The cylindrical trap is in this case to be bisected by a vertical rather than a horizontal plane, giving 
\begin{equation}
t_w = \frac{\pi r}{v}.
\end{equation}
This then yields
\begin{equation}
\label{eqn:T2dBzdx}
T_{2,\mathrm{hgi}}\sim \frac{9\pi v}{ 8 r^3 \gamma_n^2 (\partial B_z/\partial x)^2},
\end{equation}
where the subscript ``hgi'' indicates that this contribution arises from intrinsic depolarization due to the horizontal gradient.

Once again, this result (lower solid blue line in \figabbr \ref{fig:T2}) provides a very resonable approximation to our simulations (red triangles -- upwards-pointing, diffuse; downwards-pointing, 80\% specular).  The $\sim$1.5 orders of magnitude difference in response between the vertical and horizontal gradients arises principally because the bottle is four times wider than it is tall, and the respective dimensions enter to the third power.     \figabbr \ref{fig:deltah_shift} shows (green dot-dashed line) $\alpha$ as a function of the applied horizontal gradient.   

Within our apparatus, it is difficult to tune the horizontal gradient in $B_0$ to better than about 30\,pT/cm, corresponding to  1\,nT (one part per thousand) difference from one side of the bottle to the other. We note that 50\,pT/cm would yield a $T_2$ of 700\,s, perfectly consistent with that typically observed in the actual experiment and able to explain the reduction from an initial polarization of $\alpha = 0.86$ when the trap is first filled to $\alpha = 0.67$ at 220\,s storage time in the absence of a vertical gradient. 

We can also calculate, just as we did for the vertical gradient, the fractional decrease in the frequency ratio $R$ that we might expect to see as a result of the spectral dependence of the intrinsic depolarization in this horizontal magnetic-field gradient.  This is shown as a red dashed line in \figabbr \ref{fig:deltah_shift}.  We see that at  horizontal gradients of around 50\,pT/cm, the slope of $R$ vs.\ $\partial B_z/\partial z$ (\eqnabbr \ref{eqn:deltaR}) decreases by about 10\%.  This factor would be a constant, independent of the applied vertical gradient -- it would not impose any curvature upon the vertical-gradient dependence.  However, these calculations are for illustration only: we remind the reader that the frequency averaging implicit here is invalid when using the Ramsey resonance technique.

\subsection{Horizontal fields: $\partial B_t/\partial x$}

We consider here additional weak fields $B_t$ that are everywhere parallel to the $xy$ plane. If uniform, such fields simply act to produce a small tilt in the net direction of the main holding field  $B_0$, and -- since the perpendicular components add quadratically -- a tiny change in its magnitude.  The resulting field would still be uniform, leaving both the depolarization rate and the frequency ratio $R$ unaltered.  

Such horizontal fields may of course have gradients of their own, e.g.\ if they are quadrupole-type fields of the form $B_x = qy$, $B_y = qx$, as discussed in \secabbr VI.c of \cite{Pendlebury_2004}.   The direction of the total field will alter slightly from one side of  the bottle to the other, but the UCN spins follow these changes adiabatically during their trajectory.  

To understand the process in simple terms, let us first consider a neutron polarized with its spin along the $\hat{z}$ axis.    If the cell has a difference $\Delta B_t$ in a transverse (i.e.\ horizontal; $x$ or $y$) field component from one side to the other, then on traversing the cell the UCN sees the $\vec{B}$ field tilt through an angle $\phi = \Delta B_t/B_z$ in a time $t_c = 2r/v_t$, where $v_t$ is the relevant transverse velocity.  The angular frequency of this tilting motion is therefore $\omega_\mathrm{tilt} = (\Delta B_t/B_z)\cdot(v_t/2r)$.  To keep $\vec{B}$ steady, we go to a reference frame rotating at $\omega_f = \omega_\mathrm{tilt}$.  To see the correct spin motion in this frame, we have to add the field
\begin{equation}
\label{eqn:Bprime}
B^\prime_{t} = \frac{1}{\gamma_n}\omega_f = \frac{\Delta B_t}{B_z}\frac{v_t}{2 \gamma_n r}. 
\end{equation}
A new $B^\prime_{t}$ must be used after each wall collision, since $\omega_f$ changes abruptly at that point. The result is that the spin of any one UCN executes a random walk, tracing out cones of small opening angles $\theta_1, \theta_2, \theta_3 ...$, where $\theta = B^\prime_{t}/B_z$, in the vicinity of the $\hat{z}$ direction.  Assuming $N$ such wall collisions during a storage time $t = Nt_c$, these small angles add vectorially to give a total angular displacement of magnitude
\begin{equation}
\theta_t = \sqrt{N}\langle \theta \rangle = \sqrt{t\frac{v_t}{2r}}\frac{B^\prime_{t}}{B_z}.
\end{equation}
Following the same methodology as for \eqnabbr \ref{eqn:T2dBzdz}, and substituting for $B^\prime_t$ from \eqnabbr \ref{eqn:Bprime}, we arrive at 
\begin{equation}
\label{eqn:T1_jmp}
T_{1,\mathrm{tfi}} \sim \frac{80 r^3 \gamma_n^2}{v^3} \frac{ B_z^4} {\Delta B_t^2},
\end{equation}
where we now refer to the longitudinal spin-relaxation time $T_1$ rather than $T_2$ because we began with the spin aligned along $\hat{z}$.  The subscript ``tfi'' refers to ``transverse-field, intrinsic'', and we have taken $v_t^2 = v^2/3$.  This derivation is of course extremely simplistic (for example, use of the mean free path $\lambda$ rather than the cell diameter would immediately reduce $T_{1,\mathrm{tfi}}$ by a factor $\sim 3$ for our trap geometry).  However, it gives interesting insight, and  (given that we are in the ``high-field'' regime where the spin-precession frequency is substantially higher than the collision frequency) we note that its dependence upon parameters is identical to that of the rather more sophisticated  \eqnabbr 66  in \cite{Kleppner_1962}. 

This particular depolarization mechanism is less effective by a factor of two when acting upon a spin precessing in the horizontal plane rather than aligned with $\hat{z}$, for the simple reason that $B_x$ components do not affect the $x$ component of spin, and similarly for $y$ components.  This leads to the well known result for this case \cite{McGregor_1990}
\begin{equation}
T_{2,\mathrm{tfi}} = 2T_{1,\mathrm{tfi}}.
\end{equation}

Putting in realistic numbers for our apparatus (a few nT for $\Delta B_t$, and $v \sim 2$\,m/s), we find that \eqnabbr \ref{eqn:T1_jmp} predicts $T_{1,\mathrm{tfi}}$ (and therefore $T_{2,\mathrm{tfi}}$) values of order 10$^6$ seconds.  We certainly cannot expect to be sensitive to this.  In any case, we have observed $T_1$ times in excess of 1000 seconds, implying that $T_{2,\mathrm{tfi}}$ > 2000 seconds, so this is clearly not a dominant effect.

In terms of frequency shifts, the mercury atoms will average out the horizontal components, whereas the neutrons remain sensitive to the total field magnitude.  This gives rise to a change in the frequency ratio $R$ of 
\begin{equation}
\frac{\delta R}{R} = \frac{q^2 r^2}{4B_0^2},
\end{equation}
where, as before, the radius of the trap is $r$.  This will be a constant shift, independent of the applied vertical gradient.  In the case of the EDM spectrometer, where horizontal field components are several hundred to a thousand times smaller than the vertical field, the resulting frequency shifts are of the order of a part per million or less.

\subsection{Wall collisions}

The cell walls may contain tiny magnetic impurities.  Collisions with these would disturb the spins on timescales much shorter than the Larmor precession period.  Since such perturbations can affect any orientation of spin equally, one can anticipate that 
\begin{equation}
T_{2,\mathrm{wall}} = T_{1,\mathrm{wall}}.
\end{equation}
As noted above, we have measured $T_1$ to be in excess of 1000\,s, which therefore sets a lower limit of 1000\,s on the contribution to $T_2$ arising from wall collisions.  This is therefore unlikely to be a significant source of depolarization.

\section{Polarization vs.\ applied vertical gradient}
\label{sec:alpha_peak}

Having established the expected response to the intrinsic depolarization, we now go on to look at the effects of the gravitationally enhanced depolarization, using calculations as discussed in \secabbr \ref{sec:calcs} above.

\begin{figure} [ht]
\begin{center}
\resizebox{0.5\textwidth}{!}{
\includegraphics{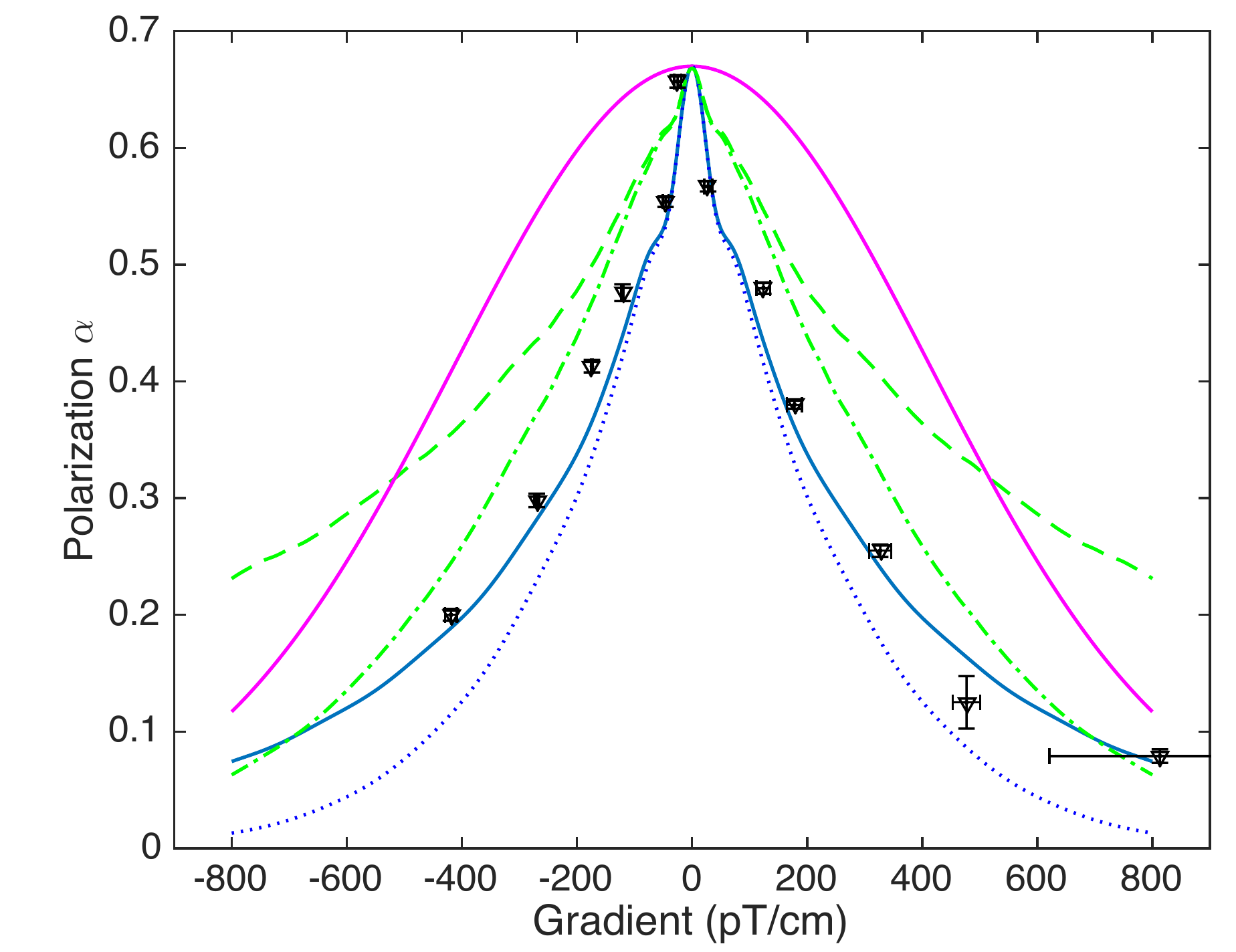}
}
\end{center}
\caption{(Color online) Depolarization as a function of applied vertical magnetic-field gradient.  Light magenta line: calculated intrinsic contribution.  Blue solid and green dashed lines are the enhanced contributions applied to the SE and MC spectra, respectively.  The blue dotted and green dot-dashed lines show the combined effects.  Measured data points are represented as black triangles.}
\label{fig:alpha_peak}
\end{figure}

We show in \figabbr \ref{fig:alpha_peak} the residual polarization, after 180\,s of storage, as a function of the applied vertical magnetic-field gradient.  The  data points (black triangles) represent measurements \cite{Afach_2014} made with the magnetic holding field $B_0$ pointing downwards.   These measurements were made in 2012, more than two years before the spin-echo measurements of the UCN spectrum, but we have no reason to suspect  that the spectrum would have changed during the intervening period.  The solid line in light magenta shows the approximate expected contribution of the intrinsic depolarization, based on the formulae of \eqnabbr \ref{eqn:T2dBzdz} and \eqnabbr \ref{eqn:T2dBzdx}; its profile should be correct, although there is uncertainty over its scale because we do not know the extent to which reflection of UCN within the trap is specular.   The blue solid (green dashed) line shows the contribution of gravitationally enhanced depolarization, using the measured SE (MC) spectrum as input.  The blue dotted (green dot-dashed) line shows the combined calculated effect.   In each case the calculated profiles are normalized to reflect the peak measured value of 0.67, which is a result both of imperfect initial polarization and also of intrinsic depolarization e.g.\ from horizontal field gradients of the form $\partial B_z/\partial x$.  $B_0$ up data are omitted from this plot for clarity; they are similar in form to the data shown, but with a somewhat lower maximum value of about 55\%.

There are two striking features about this plot.  The first is the very distinctive peaked shape of the profile near the maximum.  The intrinsic depolarization mechanism has a very soft peak, parabolic in nature.  In contrast, the gravitationally enhanced component is almost triangular in form, precisely mirroring the behaviour of the data.

The second feature of interest is the close match in the polarization profile across a wide range of gradients.  No parameters were optimized in the calculated curves beyond the normalization of the peak value (equivalent to assuming $\partial B_z/\partial x$ = 50\,pT/cm).  As noted above, the MC spectrum is expected to be a little too hard, and we see that the data lie below the corresponding (green) lines as one would anticipate.  The measured SE spectrum, on the other hand, is known to be a little too soft, since it is representative of a 220\,s storage time whereas the data points were measured at 180\,s.  One would therefore expect the combined effects of intrinsic and enhanced contributions (dotted blue)  to lie a little below the data points, as indeed they do.  It would appear that, rather fortuitously, the offset from the use of a softened spectrum is here almost exactly compensated  by the additional contribution of the intrinsic depolarization, leaving the calculated enhanced contribution more or less perfectly aligned with the data.

\section{Frequency shifts at large vertical field gradients}
\label{sec:R_curve_large}

We now turn to measurements of the ratio of neutron to mercury precession frequencies under applied vertical magnetic-field gradients. 

\begin{figure} [ht]
\begin{center}
\resizebox{0.5\textwidth}{!}{
\includegraphics{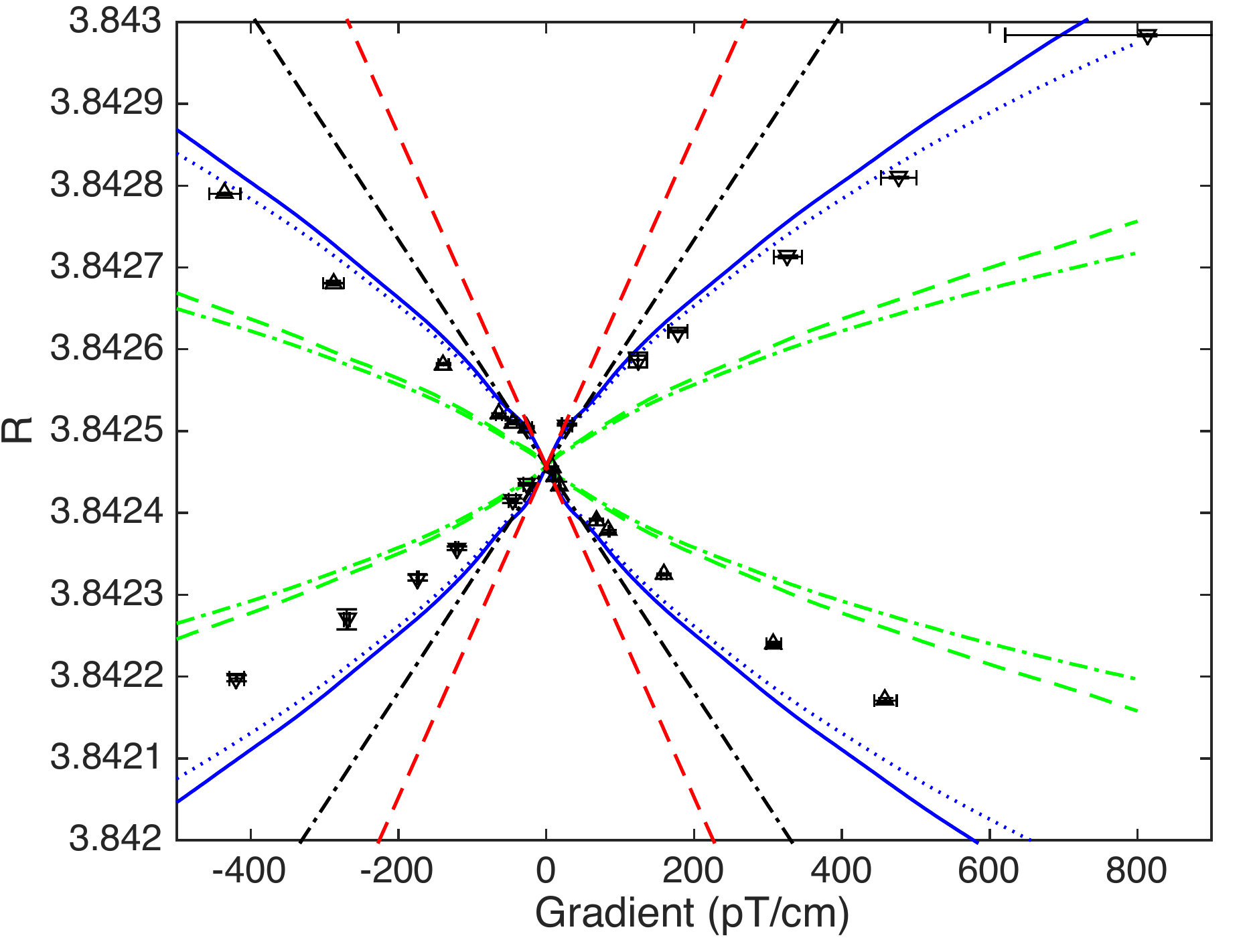}
}
\end{center}
\caption{(Color online) Ratio of neutron to mercury frequencies as a function of the applied vertical magnetic-field gradient. Triangular upwards-pointing data points (running downwards diagonally) are for $B_0$ up; triangular downwards-pointing data points (running upwards diagonally) are for $B_0$ down.  Red dashed lines show the expectation from \eqnabbr \ref{eqn:deltaR}, using $\Delta h$ from the SE spectrum.  The black dot-dashed lines represent a fit to data in the central region, with gradients of less than 60\,pT/cm.  The blue solid (green dashed) line includes gravitationally enhanced depolarization, based on the SE (MC) spectrum, with the adjacent blue dotted (green dot-dashed) lines including the effect of intrinsic depolarization.
 }
\label{fig:R_curves_large}
\end{figure}

\figabbr \ref{fig:R_curves_large} shows the measured data alongside the results from calculations of the effect of gravitationally enhanced depolarization, based upon the  measured SE (solid blue line) and simulated MC (dashed green line) spectra discussed in \secabbr \ref{sec:spectra} above.  The adjacent blue dotted and green dot-dashed lines include the approximate respective contributions from intrinsic depolarization, with the sines and cosines of the contributing phases (see \eqnabbr \ref{eqn:ref_phase}) weighted as $\alpha^2$.  As anticipated, the intrinsic depolarization has little additional effect.

Both the SE and the MC spectra result in the right general trend, i.e.\ curvature of the appropriate form.  However, as one might expect, the (stiffer) simulated spectrum results in curves that are less steep than the data, whereas the (softer) spin-echo spectrum, with its larger $\Delta h$,  results in curves that are rather steeper.  

Also shown in this figure as a pair of red dashed lines is the expected response based on \eqnabbr \ref{eqn:deltaR}, using the $\Delta h$ from the SE spectrum (with no depolarization), as well as (black dot-dashed lines) a fit to  the data points with vertical gradients of less than 60\,pT/cm.  Bearing in mind both that the SE spectrum is a little too soft and that no depolarization effects at all are included, one would anticipate that the former lines would be slightly steeper than the latter, as is indeed the case.

\section{Frequency shifts at small vertical field gradients}
\label{sec:R_curve_closeup}

\begin{figure} [ht]
\begin{center}
\resizebox{0.5\textwidth}{!}{
\includegraphics{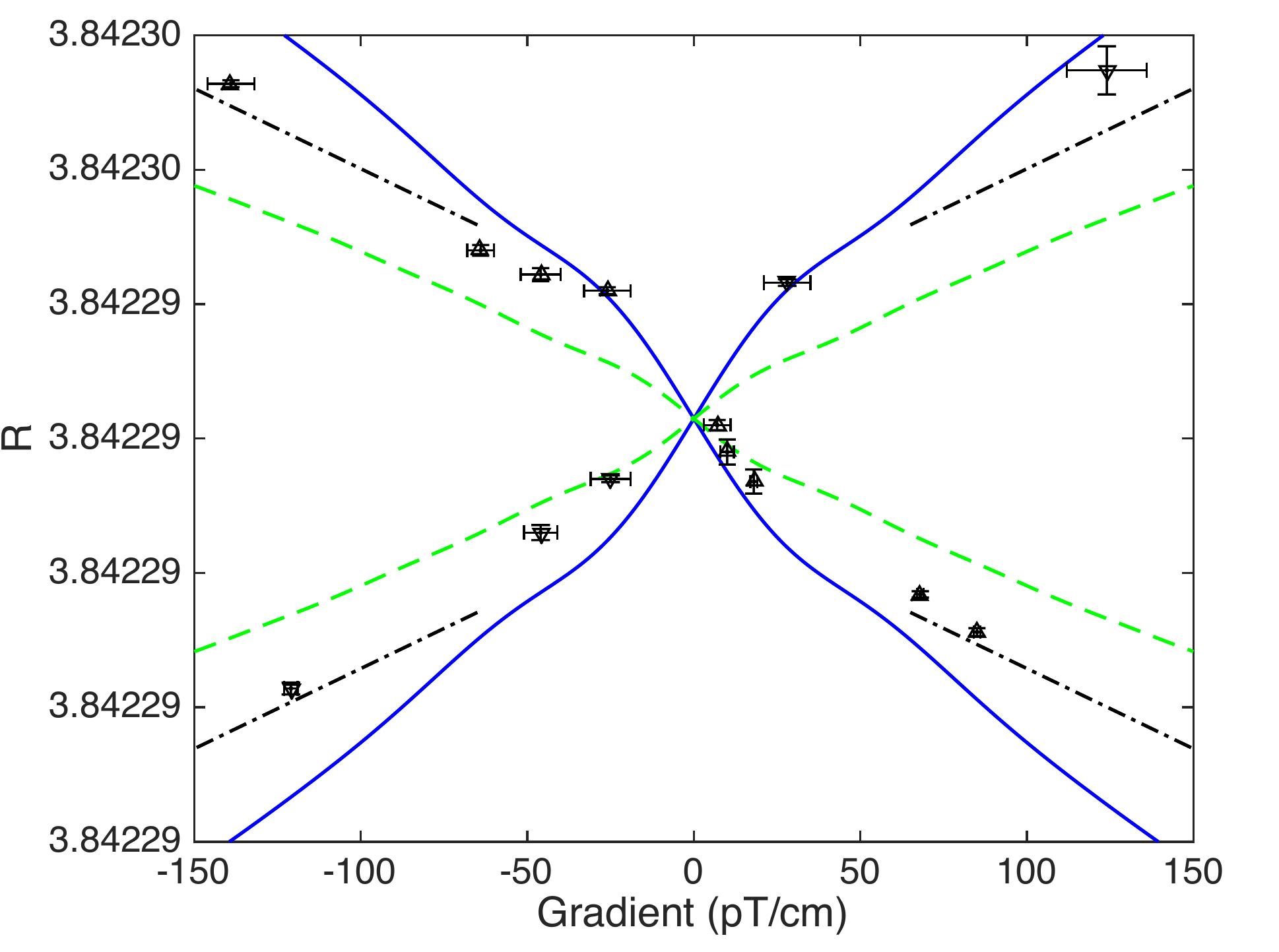}
}
\end{center}
\caption{(Color online) A closeup of the central region of the curves showing the dependence of the ratio of neutron to mercury frequencies upon the applied vertical magnetic-field gradient.  Triangular upwards- and downwards-pointing data markers once again represent measurements taken with the holding field $B_0$ aligned vertically upwards and downwards respectively.  Solid blue and dashed green lines again arise from calculations based on the spin-echo and simulated spectra, respectively.  The black dot-dashed lines in the corner regions represent the trend of the data at higher gradients. 
 }
\label{fig:R_curves_central}
\end{figure}

The effect that we wish to discuss here is arguably more subtle.  We focus upon the very central region of the frequency-ratio curves, as shown in \figabbr \ref{fig:R_curves_central}.  We again show (via the blue solid and green dashed lines, respectively) the results of calculations of gravitationally enhanced depolarization based upon the SE and MC spectra.

It is apparent that there is a change in the slope of the lines as the crossing point is approached.  It is visible in the data, where the trend at higher gradients is highlighted by the black dot-dashed lines: these represent a common fit of all of the data with gradients of more than 100\,pT/cm to the function
\begin{equation}
\left| R-R_0\right| = m\left|\frac{\partial B_z}{\partial z}\right| +c,
\end{equation}
where $R_0 = 3.8424574$ is the crossing-point value of $R$ \cite{Afach_2014}.  If extrapolated to lower gradients, these lines would clearly result in a significant discontinuity.

We ascribe this phenomenon to Ramsey wrapping, which makes the spreading low-energy tail of the array of integrated phases indistinguishable from the contributions of high-energy UCN.  This moderates the frequency shift  as described in \cite{Harris_2014} and shown in \figabbr 4 therein.  Intuitively, we would expect this to start happening when the difference in magnetic field between the bottom of the bottle (where the low-energy UCN preferentially spend their time) and the center of the bottle (which is the average position for higher-energy neutrons) is sufficient to produce a phase shift of between $\pi$ and 2$\pi$ over the 180\,s storage time.  This amounts to about 100-200\,pT over the 6\,cm half-height, i.e.\ a gradient of about 15-30\,pT/cm, which is precisely where we observe it happening.  

The particularly keen-eyed reader may be able to perceive a further slight curvature in the calculated curve at a gradient of about $\pm 50$\,pT/cm.  This is due to that same low-energy tail wrapping itself around for a second time.   The data do not have adequate resolution to discern this effect.  

We note finally that  both the curves and the real physical behavior of the system are expected to be symmetric with respect to the crossing point (indeed, the calculations represented here were generated in a single quadrant only, and then reflected through $\partial B_z/\partial z$ = 0 and through $R = R_0$). Therefore, data taken at points
using the same positive and negative gradients and with both $B_0$ field directions would allow one to extract the correct crossing point without
knowing the curvature. This is an important safeguard for future nEDM data taking.

\section{Implications for nEDM}
\label{sec:edm_implications}

The leading systematic error in the nEDM measurement described in \cite{Baker_2006} arose from shifts in the Larmor precession frequency brought about by the interplay between (a) small magnetic-field gradients within the apparatus and (b) the motional magnetic fields due to the particles (both UCN and, in particular, the mercury atoms used for magnetometry) moving through the applied electric field \cite{Pendlebury_2004}.  This effect was compensated by considering the behaviour of the observed apparent EDM signal over a range of magnetic-field gradients. There was no direct measurement of the field gradient: instead,  it was parameterized (\eqnabbr \ref{eqn:deltaR}) by the ratio $R$ of neutron to mercury precession frequencies.  It is now clear, however, that even at quite moderate gradients $R$ is subject to the nonlinearities discussed above.  Furthermore, it is stated in \cite{Baker_2006} that the height difference $\Delta h$ between the UCN and the mercury atoms was 2.8 mm, with a precision of 4\%.  This latter was based upon measurements of the frequency response within a variable-height trap; in fact those measurements also would have been affected by  gravitationally enhanced depolarization.  The UCN energy spectrum was undoubtedly softer than had been thought, implying that the calculated slope of the lines in \figabbr 2 of \cite{Baker_2006} was steeper than it should have been.  The relatively good match of the calculated to the actual slope of the fitted line is due in part to the nonlinear nature of the dependence of $R$ upon $\partial B_z/\partial z$, which would be similar in form to that shown in \figabbr \ref{fig:R_curves_central} above and in \figabbr 4 of \cite{Harris_2014}, and which would result in a steeper-than-expected slope once beyond the linear region.  

Since the data were taken more or less symmetrically about the crossing point, this effect is unlikely to produce a very substantial change in the nEDM limit in this case.  Nonetheless, a detailed reanalysis is now underway, and it is expected that a revised result will emerge shortly.  Future measurements will doubtless enjoy the advantage of better diagnostics both of the magnetic field (with improved magnetometry) and of the energy spectrum (using the spin-echo technique).

\section{Conclusions}

Measurements undertaken at the EDM spectrometer at PSI, showing the dependence upon applied vertical magnetic-field gradients of depolarization rates and of the neutron precession-frequency, have clearly demonstrated features that are characteristic of the anticipated behaviour resulting from gravitationally enhanced depolarization and Ramsey wrapping: namely, a sharply-peaked rather than parabolic depolarization profile, and significant nonlinearities in the frequency-response curve.  Using estimates of the spectrum of stored UCN, based upon measurements using the spin-echo technique and also upon detailed simulations, we have demonstrated excellent qualitative agreement between measurements and theoretical expectations.   It also seems clear that intrinsic depolarization processes have only a marginal effect upon frequency shifts in the presence of magnetic-field gradients, and that such shifts are dominated by the gravitationally enhanced component.  

There are obvious implications for nEDM measurements, including for the analysis that led to the current world limit \cite{Baker_2006}, since the frequency-response curve is used to correct for systematic effects.

\section*{Acknowledgements}

We would like to thank the PSI staff, in particular F.~Burri and M.~Meier, for their outstanding support.  We also gratefully acknowledge the important work carried out by the mechanical workshop of the University of Fribourg Physics Department. This research was financed in part by the Fund for Scientific Research, Flanders; grant GO A/2010/10 of KU~Leuven; the Swiss National Science Foundation Projects 200020-144473 (PSI), 200021-126562 (PSI), 200020-149211 (ETH) and 200020-140421 (Fribourg); and grants ST/K001329/1, ST/M003426/1 and ST/L006472/1 from the UK's Science and Technology Facilities Council (STFC).  One of us (EW) is a PhD Fellow of the Research Foundation - Flanders (FWO). The original apparatus was funded by grants from the UK's PPARC (now STFC). The LPC Caen and the LPSC acknowledge the support of the French Agence Nationale de la Recherche (ANR) under reference ANR-09-BLAN-0046. Polish partners wish to acknowledge support from the PL-Grid infrastructure and from the National Science Centre, Poland, under grant no.\ UMO-2012/04/M/ST2/00556.

\bibliography{2015_neutron_EDM}

\begin{thebibliography}{28}
\expandafter\ifx\csname natexlab\endcsname\relax\def\natexlab#1{#1}\fi
\expandafter\ifx\csname bibnamefont\endcsname\relax
  \def\bibnamefont#1{#1}\fi
\expandafter\ifx\csname bibfnamefont\endcsname\relax
  \def\bibfnamefont#1{#1}\fi
\expandafter\ifx\csname citenamefont\endcsname\relax
  \def\citenamefont#1{#1}\fi
\expandafter\ifx\csname url\endcsname\relax
  \def\url#1{\texttt{#1}}\fi
\expandafter\ifx\csname urlprefix\endcsname\relax\def\urlprefix{URL }\fi
\providecommand{\bibinfo}[2]{#2}
\providecommand{\eprint}[2][]{\url{#2}}

\bibitem[{\citenamefont{Harris et~al.}(2014)\citenamefont{Harris, Pendlebury,
  and Devenish}}]{Harris_2014}
\bibinfo{author}{\bibfnamefont{P.~G.} \bibnamefont{Harris}},
  \bibinfo{author}{\bibfnamefont{J.~M.} \bibnamefont{Pendlebury}},
  \bibnamefont{and} \bibinfo{author}{\bibfnamefont{N.~E.}
  \bibnamefont{Devenish}}, \bibinfo{journal}{Physical Review D}
  \textbf{\bibinfo{volume}{89}} (\bibinfo{year}{2014}),
  \urlprefix\url{http://dx.doi.org/10.1103/physrevd.89.016011}.

\bibitem[{\citenamefont{Knecht}(2009)}]{knecht09}
\bibinfo{author}{\bibfnamefont{A.}~\bibnamefont{Knecht}}, Ph.D. thesis,
  \bibinfo{school}{Universit{\"a}t Z{\"u}rich} (\bibinfo{year}{2009}).

\bibitem[{\citenamefont{Pendlebury et~al.}(2004)\citenamefont{Pendlebury, Heil,
  Sobolev, Harris, Richardson, Baskin, Doyle, Geltenbort, Green, van~der
  Grinten et~al.}}]{Pendlebury_2004}
\bibinfo{author}{\bibfnamefont{J.~M.} \bibnamefont{Pendlebury}},
  \bibinfo{author}{\bibfnamefont{W.}~\bibnamefont{Heil}},
  \bibinfo{author}{\bibfnamefont{Y.}~\bibnamefont{Sobolev}},
  \bibinfo{author}{\bibfnamefont{P.~G.} \bibnamefont{Harris}},
  \bibinfo{author}{\bibfnamefont{J.~D.} \bibnamefont{Richardson}},
  \bibinfo{author}{\bibfnamefont{R.~J.} \bibnamefont{Baskin}},
  \bibinfo{author}{\bibfnamefont{D.~D.} \bibnamefont{Doyle}},
  \bibinfo{author}{\bibfnamefont{P.}~\bibnamefont{Geltenbort}},
  \bibinfo{author}{\bibfnamefont{K.}~\bibnamefont{Green}},
  \bibinfo{author}{\bibfnamefont{M.~G.~D.} \bibnamefont{van~der Grinten}},
  \bibnamefont{et~al.}, \bibinfo{journal}{Physical Review A}
  \textbf{\bibinfo{volume}{70}} (\bibinfo{year}{2004}),
  \urlprefix\url{http://dx.doi.org/10.1103/physreva.70.032102}.

\bibitem[{\citenamefont{Baker et~al.}(2006)\citenamefont{Baker, Doyle,
  Geltenbort, Green, van~der Grinten, Harris, Iaydjiev, Ivanov, May, Pendlebury
  et~al.}}]{Baker_2006}
\bibinfo{author}{\bibfnamefont{C.~A.} \bibnamefont{Baker}},
  \bibinfo{author}{\bibfnamefont{D.~D.} \bibnamefont{Doyle}},
  \bibinfo{author}{\bibfnamefont{P.}~\bibnamefont{Geltenbort}},
  \bibinfo{author}{\bibfnamefont{K.}~\bibnamefont{Green}},
  \bibinfo{author}{\bibfnamefont{M.~G.~D.} \bibnamefont{van~der Grinten}},
  \bibinfo{author}{\bibfnamefont{P.~G.} \bibnamefont{Harris}},
  \bibinfo{author}{\bibfnamefont{P.}~\bibnamefont{Iaydjiev}},
  \bibinfo{author}{\bibfnamefont{S.~N.} \bibnamefont{Ivanov}},
  \bibinfo{author}{\bibfnamefont{D.~J.~R.} \bibnamefont{May}},
  \bibinfo{author}{\bibfnamefont{J.~M.} \bibnamefont{Pendlebury}},
  \bibnamefont{et~al.}, \bibinfo{journal}{Phys. Rev. Lett.}
  \textbf{\bibinfo{volume}{97}} (\bibinfo{year}{2006}),
  \urlprefix\url{http://dx.doi.org/10.1103/physrevlett.97.131801}.

\bibitem[{\citenamefont{Afach et~al.}(2015{\natexlab{a}})\citenamefont{Afach,
  Baker, Ban et~al.}}]{afach15c}
\bibinfo{author}{\bibfnamefont{S.}~\bibnamefont{Afach}},
  \bibinfo{author}{\bibfnamefont{C.}~\bibnamefont{Baker}},
  \bibinfo{author}{\bibfnamefont{G.}~\bibnamefont{Ban}}, \bibnamefont{et~al.},
  \bibinfo{journal}{Submitted to European Physical Journal D}
  (\bibinfo{year}{2015}{\natexlab{a}}),
  \bibinfo{note}{http://arxiv.org/abs/1503.08651}.

\bibitem[{\citenamefont{Baker et~al.}(2011)\citenamefont{Baker, Ban, Bodek,
  Burghoff, Chowdhuri, Daum, Fertl, Franke, Geltenbort, Green
  et~al.}}]{Baker_2011}
\bibinfo{author}{\bibfnamefont{C.}~\bibnamefont{Baker}},
  \bibinfo{author}{\bibfnamefont{G.}~\bibnamefont{Ban}},
  \bibinfo{author}{\bibfnamefont{K.}~\bibnamefont{Bodek}},
  \bibinfo{author}{\bibfnamefont{M.}~\bibnamefont{Burghoff}},
  \bibinfo{author}{\bibfnamefont{Z.}~\bibnamefont{Chowdhuri}},
  \bibinfo{author}{\bibfnamefont{M.}~\bibnamefont{Daum}},
  \bibinfo{author}{\bibfnamefont{M.}~\bibnamefont{Fertl}},
  \bibinfo{author}{\bibfnamefont{B.}~\bibnamefont{Franke}},
  \bibinfo{author}{\bibfnamefont{P.}~\bibnamefont{Geltenbort}},
  \bibinfo{author}{\bibfnamefont{K.}~\bibnamefont{Green}},
  \bibnamefont{et~al.}, \bibinfo{journal}{Physics Procedia}
  \textbf{\bibinfo{volume}{17}}, \bibinfo{pages}{159} (\bibinfo{year}{2011}),
  \urlprefix\url{http://dx.doi.org/10.1016/j.phpro.2011.06.032}.

\bibitem[{\citenamefont{Lauss}(2014)}]{Lauss_2014}
\bibinfo{author}{\bibfnamefont{B.}~\bibnamefont{Lauss}},
  \bibinfo{journal}{Physics Procedia} \textbf{\bibinfo{volume}{51}},
  \bibinfo{pages}{98} (\bibinfo{year}{2014}),
  \urlprefix\url{http://dx.doi.org/10.1016/j.phpro.2013.12.022}.

\bibitem[{\citenamefont{Afach et~al.}(2014{\natexlab{a}})\citenamefont{Afach,
  Baker, Ban, Bison, Bodek, Burghoff, Chowdhuri, Daum, Fertl, Franke
  et~al.}}]{Afach_2014}
\bibinfo{author}{\bibfnamefont{S.}~\bibnamefont{Afach}},
  \bibinfo{author}{\bibfnamefont{C.}~\bibnamefont{Baker}},
  \bibinfo{author}{\bibfnamefont{G.}~\bibnamefont{Ban}},
  \bibinfo{author}{\bibfnamefont{G.}~\bibnamefont{Bison}},
  \bibinfo{author}{\bibfnamefont{K.}~\bibnamefont{Bodek}},
  \bibinfo{author}{\bibfnamefont{M.}~\bibnamefont{Burghoff}},
  \bibinfo{author}{\bibfnamefont{Z.}~\bibnamefont{Chowdhuri}},
  \bibinfo{author}{\bibfnamefont{M.}~\bibnamefont{Daum}},
  \bibinfo{author}{\bibfnamefont{M.}~\bibnamefont{Fertl}},
  \bibinfo{author}{\bibfnamefont{B.}~\bibnamefont{Franke}},
  \bibnamefont{et~al.}, \bibinfo{journal}{Physics Letters B}
  \textbf{\bibinfo{volume}{739}}, \bibinfo{pages}{128}
  (\bibinfo{year}{2014}{\natexlab{a}}),
  \urlprefix\url{http://dx.doi.org/10.1016/j.physletb.2014.10.046}.

\bibitem[{\citenamefont{Baker et~al.}(2014)\citenamefont{Baker, Chibane,
  Chouder, Geltenbort, Green, Harris, Heckel, Iaydjiev, Ivanov, Kilvington
  et~al.}}]{Baker_2014}
\bibinfo{author}{\bibfnamefont{C.}~\bibnamefont{Baker}},
  \bibinfo{author}{\bibfnamefont{Y.}~\bibnamefont{Chibane}},
  \bibinfo{author}{\bibfnamefont{M.}~\bibnamefont{Chouder}},
  \bibinfo{author}{\bibfnamefont{P.}~\bibnamefont{Geltenbort}},
  \bibinfo{author}{\bibfnamefont{K.}~\bibnamefont{Green}},
  \bibinfo{author}{\bibfnamefont{P.}~\bibnamefont{Harris}},
  \bibinfo{author}{\bibfnamefont{B.}~\bibnamefont{Heckel}},
  \bibinfo{author}{\bibfnamefont{P.}~\bibnamefont{Iaydjiev}},
  \bibinfo{author}{\bibfnamefont{S.}~\bibnamefont{Ivanov}},
  \bibinfo{author}{\bibfnamefont{I.}~\bibnamefont{Kilvington}},
  \bibnamefont{et~al.}, \bibinfo{journal}{Nuclear Instruments and Methods in
  Physics Research Section A: Accelerators, Spectrometers, Detectors and
  Associated Equipment} \textbf{\bibinfo{volume}{736}}, \bibinfo{pages}{184}
  (\bibinfo{year}{2014}),
  \urlprefix\url{http://dx.doi.org/10.1016/j.nima.2013.10.005}.

\bibitem[{\citenamefont{Knowles et~al.}(2009)\citenamefont{Knowles, Bison,
  Castagna, Hofer, Mtchedlishvili, Pazgalev, and Weis}}]{Knowles_2009}
\bibinfo{author}{\bibfnamefont{P.}~\bibnamefont{Knowles}},
  \bibinfo{author}{\bibfnamefont{G.}~\bibnamefont{Bison}},
  \bibinfo{author}{\bibfnamefont{N.}~\bibnamefont{Castagna}},
  \bibinfo{author}{\bibfnamefont{A.}~\bibnamefont{Hofer}},
  \bibinfo{author}{\bibfnamefont{A.}~\bibnamefont{Mtchedlishvili}},
  \bibinfo{author}{\bibfnamefont{A.}~\bibnamefont{Pazgalev}}, \bibnamefont{and}
  \bibinfo{author}{\bibfnamefont{A.}~\bibnamefont{Weis}},
  \bibinfo{journal}{Nuclear Instruments and Methods in Physics Research Section
  A: Accelerators, Spectrometers, Detectors and Associated Equipment}
  \textbf{\bibinfo{volume}{611}}, \bibinfo{pages}{306} (\bibinfo{year}{2009}),
  \urlprefix\url{http://dx.doi.org/10.1016/j.nima.2009.07.079}.

\bibitem[{\citenamefont{Afach et~al.}(2015{\natexlab{b}})\citenamefont{Afach,
  Ban et~al.}}]{afach15}
\bibinfo{author}{\bibfnamefont{S.}~\bibnamefont{Afach}},
  \bibinfo{author}{\bibfnamefont{G.}~\bibnamefont{Ban}}, \bibnamefont{et~al.},
  \bibinfo{journal}{Submitted to European Physical Journal A}
  (\bibinfo{year}{2015}{\natexlab{b}}),
  \bibinfo{note}{http://arxiv.org/abs/1502.06876}.

\bibitem[{\citenamefont{Afach et~al.}(2014{\natexlab{b}})\citenamefont{Afach,
  Bison, Bodek, Burri, Chowdhuri, Daum, Fertl, Franke, Grujic, H{\'{e}}laine
  et~al.}}]{Afach_2014b}
\bibinfo{author}{\bibfnamefont{S.}~\bibnamefont{Afach}},
  \bibinfo{author}{\bibfnamefont{G.}~\bibnamefont{Bison}},
  \bibinfo{author}{\bibfnamefont{K.}~\bibnamefont{Bodek}},
  \bibinfo{author}{\bibfnamefont{F.}~\bibnamefont{Burri}},
  \bibinfo{author}{\bibfnamefont{Z.}~\bibnamefont{Chowdhuri}},
  \bibinfo{author}{\bibfnamefont{M.}~\bibnamefont{Daum}},
  \bibinfo{author}{\bibfnamefont{M.}~\bibnamefont{Fertl}},
  \bibinfo{author}{\bibfnamefont{B.}~\bibnamefont{Franke}},
  \bibinfo{author}{\bibfnamefont{Z.}~\bibnamefont{Grujic}},
  \bibinfo{author}{\bibfnamefont{V.}~\bibnamefont{H{\'{e}}laine}},
  \bibnamefont{et~al.}, \bibinfo{journal}{J. Appl. Phys.}
  \textbf{\bibinfo{volume}{116}}, \bibinfo{pages}{084510}
  (\bibinfo{year}{2014}{\natexlab{b}}),
  \urlprefix\url{http://dx.doi.org/10.1063/1.4894158}.

\bibitem[{\citenamefont{Green et~al.}(1998)\citenamefont{Green, Harris,
  Iaydjiev, May, Pendlebury, Smith, van~der Grinten, Geltenbort, and
  Ivanov}}]{Green_1998}
\bibinfo{author}{\bibfnamefont{K.}~\bibnamefont{Green}},
  \bibinfo{author}{\bibfnamefont{P.}~\bibnamefont{Harris}},
  \bibinfo{author}{\bibfnamefont{P.}~\bibnamefont{Iaydjiev}},
  \bibinfo{author}{\bibfnamefont{D.}~\bibnamefont{May}},
  \bibinfo{author}{\bibfnamefont{J.}~\bibnamefont{Pendlebury}},
  \bibinfo{author}{\bibfnamefont{K.}~\bibnamefont{Smith}},
  \bibinfo{author}{\bibfnamefont{M.}~\bibnamefont{van~der Grinten}},
  \bibinfo{author}{\bibfnamefont{P.}~\bibnamefont{Geltenbort}},
  \bibnamefont{and} \bibinfo{author}{\bibfnamefont{S.}~\bibnamefont{Ivanov}},
  \bibinfo{journal}{Nuclear Instruments and Methods in Physics Research Section
  A: Accelerators, Spectrometers, Detectors and Associated Equipment}
  \textbf{\bibinfo{volume}{404}}, \bibinfo{pages}{381} (\bibinfo{year}{1998}),
  \urlprefix\url{http://dx.doi.org/10.1016/s0168-9002(97)01121-2}.

\bibitem[{\citenamefont{Afach et~al.}(2015{\natexlab{c}})\citenamefont{Afach,
  Ayres, Ban et~al.}}]{afach15b}
\bibinfo{author}{\bibfnamefont{S.}~\bibnamefont{Afach}},
  \bibinfo{author}{\bibfnamefont{N.}~\bibnamefont{Ayres}},
  \bibinfo{author}{\bibfnamefont{G.}~\bibnamefont{Ban}}, \bibnamefont{et~al.},
  \bibinfo{journal}{Submitted to Phys.\ Rev.\ Lett.}
  (\bibinfo{year}{2015}{\natexlab{c}}),
  \bibinfo{note}{http://arxiv.org/abs/1506.00446}.

\bibitem[{\citenamefont{Bodek et~al.}(2011)\citenamefont{Bodek, Chowdhuri,
  Daum, Fertl, Franke, Gutsmiedl, Henneck, Horras, Kasprzak, Kirch
  et~al.}}]{Bodek_2011}
\bibinfo{author}{\bibfnamefont{K.}~\bibnamefont{Bodek}},
  \bibinfo{author}{\bibfnamefont{Z.}~\bibnamefont{Chowdhuri}},
  \bibinfo{author}{\bibfnamefont{M.}~\bibnamefont{Daum}},
  \bibinfo{author}{\bibfnamefont{M.}~\bibnamefont{Fertl}},
  \bibinfo{author}{\bibfnamefont{B.}~\bibnamefont{Franke}},
  \bibinfo{author}{\bibfnamefont{E.}~\bibnamefont{Gutsmiedl}},
  \bibinfo{author}{\bibfnamefont{R.}~\bibnamefont{Henneck}},
  \bibinfo{author}{\bibfnamefont{M.}~\bibnamefont{Horras}},
  \bibinfo{author}{\bibfnamefont{M.}~\bibnamefont{Kasprzak}},
  \bibinfo{author}{\bibfnamefont{K.}~\bibnamefont{Kirch}},
  \bibnamefont{et~al.}, \bibinfo{journal}{Physics Procedia}
  \textbf{\bibinfo{volume}{17}}, \bibinfo{pages}{259} (\bibinfo{year}{2011}),
  \urlprefix\url{http://dx.doi.org/10.1016/j.phpro.2011.06.046}.

\bibitem[{\citenamefont{Golub et~al.}(1991)\citenamefont{Golub, Richardson, and
  Lamoreaux}}]{golub_UCN_book}
\bibinfo{author}{\bibfnamefont{R.}~\bibnamefont{Golub}},
  \bibinfo{author}{\bibfnamefont{D.}~\bibnamefont{Richardson}},
  \bibnamefont{and}
  \bibinfo{author}{\bibfnamefont{S.}~\bibnamefont{Lamoreaux}},
  \emph{\bibinfo{title}{Ultra-Cold Neutrons}} (\bibinfo{publisher}{Adam
  Hilger}, \bibinfo{address}{Bristol}, \bibinfo{year}{1991}).

\bibitem[{\citenamefont{Ku{\'z}niak}(2008)}]{kuzniak08}
\bibinfo{author}{\bibfnamefont{M.}~\bibnamefont{Ku{\'z}niak}}, Ph.D. thesis,
  \bibinfo{school}{Jagiellonian University} (\bibinfo{year}{2008}).

\bibitem[{\citenamefont{Pendlebury and Richardson}(1994)}]{Pendlebury_1994}
\bibinfo{author}{\bibfnamefont{J.}~\bibnamefont{Pendlebury}} \bibnamefont{and}
  \bibinfo{author}{\bibfnamefont{D.}~\bibnamefont{Richardson}},
  \bibinfo{journal}{Nuclear Instruments and Methods in Physics Research Section
  A: Accelerators, Spectrometers, Detectors and Associated Equipment}
  \textbf{\bibinfo{volume}{337}}, \bibinfo{pages}{504} (\bibinfo{year}{1994}),
  \urlprefix\url{http://dx.doi.org/10.1016/0168-9002(94)91120-7}.

\bibitem[{\citenamefont{Schearer and Walters}(1965)}]{Schearer_1965}
\bibinfo{author}{\bibfnamefont{L.~D.} \bibnamefont{Schearer}} \bibnamefont{and}
  \bibinfo{author}{\bibfnamefont{G.~K.} \bibnamefont{Walters}},
  \bibinfo{journal}{Physical Review} \textbf{\bibinfo{volume}{139}},
  \bibinfo{pages}{A1398} (\bibinfo{year}{1965}),
  \urlprefix\url{http://dx.doi.org/10.1103/physrev.139.a1398}.

\bibitem[{\citenamefont{Cates et~al.}(1988{\natexlab{a}})\citenamefont{Cates,
  Schaefer, and Happer}}]{Cates_1988}
\bibinfo{author}{\bibfnamefont{G.~D.} \bibnamefont{Cates}},
  \bibinfo{author}{\bibfnamefont{S.~R.} \bibnamefont{Schaefer}},
  \bibnamefont{and} \bibinfo{author}{\bibfnamefont{W.}~\bibnamefont{Happer}},
  \bibinfo{journal}{Physical Review A} \textbf{\bibinfo{volume}{37}},
  \bibinfo{pages}{2877} (\bibinfo{year}{1988}{\natexlab{a}}),
  \urlprefix\url{http://dx.doi.org/10.1103/physreva.37.2877}.

\bibitem[{\citenamefont{Cates et~al.}(1988{\natexlab{b}})\citenamefont{Cates,
  White, Chien, Schaefer, and Happer}}]{Cates_1988b}
\bibinfo{author}{\bibfnamefont{G.~D.} \bibnamefont{Cates}},
  \bibinfo{author}{\bibfnamefont{D.~J.} \bibnamefont{White}},
  \bibinfo{author}{\bibfnamefont{T.-R.} \bibnamefont{Chien}},
  \bibinfo{author}{\bibfnamefont{S.~R.} \bibnamefont{Schaefer}},
  \bibnamefont{and} \bibinfo{author}{\bibfnamefont{W.}~\bibnamefont{Happer}},
  \bibinfo{journal}{Physical Review A} \textbf{\bibinfo{volume}{38}},
  \bibinfo{pages}{5092} (\bibinfo{year}{1988}{\natexlab{b}}),
  \urlprefix\url{http://dx.doi.org/10.1103/physreva.38.5092}.

\bibitem[{\citenamefont{McGregor}(1990)}]{McGregor_1990}
\bibinfo{author}{\bibfnamefont{D.~D.} \bibnamefont{McGregor}},
  \bibinfo{journal}{Physical Review A} \textbf{\bibinfo{volume}{41}},
  \bibinfo{pages}{2631} (\bibinfo{year}{1990}),
  \urlprefix\url{http://dx.doi.org/10.1103/physreva.41.2631}.

\bibitem[{\citenamefont{Schmid et~al.}(2008)\citenamefont{Schmid, Plaster, and
  Filippone}}]{Schmid_2008}
\bibinfo{author}{\bibfnamefont{R.}~\bibnamefont{Schmid}},
  \bibinfo{author}{\bibfnamefont{B.}~\bibnamefont{Plaster}}, \bibnamefont{and}
  \bibinfo{author}{\bibfnamefont{B.~W.} \bibnamefont{Filippone}},
  \bibinfo{journal}{Physical Review A} \textbf{\bibinfo{volume}{78}}
  (\bibinfo{year}{2008}),
  \urlprefix\url{http://dx.doi.org/10.1103/physreva.78.023401}.

\bibitem[{\citenamefont{Lamoreaux and Golub}(2005)}]{Lamoreaux_2005}
\bibinfo{author}{\bibfnamefont{S.~K.} \bibnamefont{Lamoreaux}}
  \bibnamefont{and} \bibinfo{author}{\bibfnamefont{R.}~\bibnamefont{Golub}},
  \bibinfo{journal}{Physical Review A} \textbf{\bibinfo{volume}{71}}
  (\bibinfo{year}{2005}),
  \urlprefix\url{http://dx.doi.org/10.1103/physreva.71.032104}.

\bibitem[{\citenamefont{Golub et~al.}(2011)\citenamefont{Golub, Rohm, and
  Swank}}]{Golub_2011}
\bibinfo{author}{\bibfnamefont{R.}~\bibnamefont{Golub}},
  \bibinfo{author}{\bibfnamefont{R.~M.} \bibnamefont{Rohm}}, \bibnamefont{and}
  \bibinfo{author}{\bibfnamefont{C.~M.} \bibnamefont{Swank}},
  \bibinfo{journal}{Physical Review A} \textbf{\bibinfo{volume}{83}}
  (\bibinfo{year}{2011}),
  \urlprefix\url{http://dx.doi.org/10.1103/physreva.83.023402}.

\bibitem[{\citenamefont{Pignol and Roccia}(2012)}]{Pignol_2012}
\bibinfo{author}{\bibfnamefont{G.}~\bibnamefont{Pignol}} \bibnamefont{and}
  \bibinfo{author}{\bibfnamefont{S.}~\bibnamefont{Roccia}},
  \bibinfo{journal}{Physical Review A} \textbf{\bibinfo{volume}{85}}
  (\bibinfo{year}{2012}),
  \urlprefix\url{http://dx.doi.org/10.1103/physreva.85.042105}.

\bibitem[{\citenamefont{Kleppner et~al.}(1962)\citenamefont{Kleppner,
  Goldenberg, and Ramsey}}]{Kleppner_1962}
\bibinfo{author}{\bibfnamefont{D.}~\bibnamefont{Kleppner}},
  \bibinfo{author}{\bibfnamefont{H.~M.} \bibnamefont{Goldenberg}},
  \bibnamefont{and} \bibinfo{author}{\bibfnamefont{N.~F.}
  \bibnamefont{Ramsey}}, \bibinfo{journal}{Physical Review}
  \textbf{\bibinfo{volume}{126}}, \bibinfo{pages}{603} (\bibinfo{year}{1962}),
  \urlprefix\url{http://dx.doi.org/10.1103/physrev.126.603}.

\bibitem[{\citenamefont{Pendlebury}(1985)}]{pendlebury_kinetic_theory}
\bibinfo{author}{\bibfnamefont{J.}~\bibnamefont{Pendlebury}},
  \emph{\bibinfo{title}{Kinetic Theory}} (\bibinfo{publisher}{Student
  Monographs in Physics, Adam Hilger, Bristol}, \bibinfo{year}{1985}).

\end{thebibliography}

\end{document}